\documentclass[twocolumn,english,superscriptaddress,aps]{revtex4}
\usepackage{times}
\usepackage[T1]{fontenc}
\usepackage[latin1]{inputenc}
\usepackage{graphicx}

\makeatletter

\usepackage{graphicx}

\usepackage{graphicx}

\usepackage{graphicx}

\usepackage{graphicx}

\usepackage{babel}
\makeatother
\begin{document}

\title{Lattice structure and vibrational properties of the same nano-object}

\author{Jannik C. Meyer}

\thanks{These authors contributed equally to this work}

\affiliation{Max Planck Institute for Solid State Research, Stuttgart, Germany}

\author{Matthieu Paillet}

\thanks{These authors contributed equally to this work}

\author{Jean-Louis Sauvajol}

\affiliation{Groupe de Dynamique des Phases Condensees, Universite de Montpellier
II, France}

\author{Georg S. Duesberg}

\affiliation{Infineon Technologies Corporate Research, Munich, Germany}

\author{Siegmar Roth}

\affiliation{Max Planck Institute for Solid State Research, Stuttgart, Germany}

\begin{abstract}
We present a procedure for determining independently the lattice structure
and the vibrational properties of the same individual nano-object.
For the example of an individual single-walled carbon nanotube we
demonstrate the determination of the structural indices (n,m) of the
nanotube by electron diffraction and of the frequencies of vibrational
modes by micro-Raman spectroscopy. The precise and independent determination
of both structure and mode frequencies allows for direct and unambiguous
verification of molecular dynamical calculations and of conclusions
drawn from Raman-only experiments.
\end{abstract}
\maketitle
\hyphenation{nano-tube}

Modeling the dynamics of a physical system at the atomic level requires
knowledge, or assumptions, about the lattice structure or the positions
of the individual atoms within the object. The properties of nanoscopic
objects often depend critically on the position of each atom as finite-size
and quantization effects play an important role. For carbon nanotubes,
for example, the electronic, mechanical, and vibrational properties
vary significantly depending on their structure. Therefore a comparison
of experimental data from single objects with theoretical predictions
will be directly possible if the structure of the object is known
and independently determined.

A carbon nanotube can be considered as a graphene sheet rolled into
a cylinder. The nanotube indices (n,m) describe the possible nanotube
structures for single-walled nanotubes (SWNTs) \cite{IijimaSWNT1993}.
Depending on the indices, the nanotube can be either metallic or semiconducting
with varying band-gaps. Also the characteristic features of the Raman
spectrum, like the so-called radial breathing mode (RBM), usually
in the 100-300~cm$^{-1}$ frequency range, and the so-called tangential
modes (TM), usually within 1500-1600~cm$^{-1}$, depend on the nanotube
indices \cite{DresselhausAdvPhys00}. As long as modelization is used
to derive the structural information from the spectroscopic data,
a verification of the model itself is limited. We present an independent
determination of the nanotube structure in combination with Raman
spectroscopy on the same object.

The structure, that is the indices, can be determined by electron
diffraction in a transmission electron microscope (TEM). However combining
the electron microscopic investigation with other measurements on
the same object is an experimental challenge. A few groups, including
our own, have achieved high-resolution transmission electron microscopy
and transport measurements on the same individual single-walled carbon
nanotube \cite{MeyerTEMTransport04,KasumovSWNTSupercurrent99}. A
combination with electron diffraction has been achieved only for nanotubes
with multiple walls \cite{LinkingNMandTransportDWNTwPiezo}. Transport
measurements, however, probe not only the properties of the nanotube
but also of the contacts, making the interpretation much more difficult.

Raman spectroscopy, on the other hand, directly yields information
about the vibrational properties of the investigated object. The frequencies
of the Raman-active modes of the system can be determined with high
accuracy, and can be directly compared with the results of the modelizations
of the vibrational dynamics of the investigated nano-object.

\begin{figure}
\includegraphics[%
  width=8.5cm]{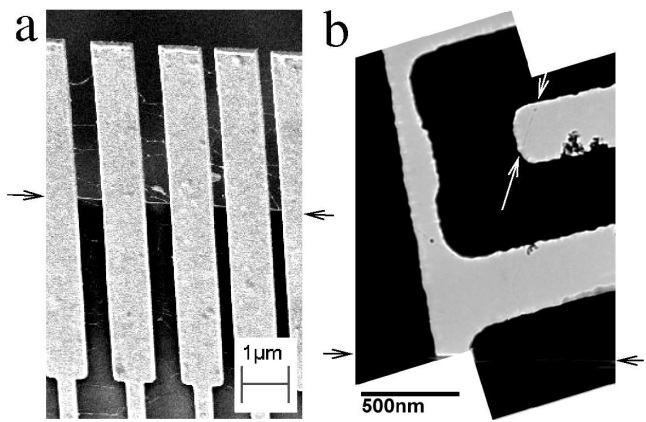}

\caption{(a) SEM image of an a sample showing carbon nanotubes suspended within
the metal structure. The dark arrows indicate the edge of the etched
substrate: The section above this line is free-standing and accessible
by TEM, while the lower part is above the remaining substrate. (b)
Low-magnification TEM Overview image of a freestanding metal structure
with suspended carbon nanotubes. The nanotube indicated by the white
arrow is examined by Raman spectroscopy, high-resolution TEM and electron
diffraction. This image is used to locate the carbon nanotube with
the optical microscope of the Raman spectrometer. \label{cap:SEM}\label{cap:Lowmag-TEM}}
\end{figure}

We show a simple {[}but completely new{]} procedure to create arbitrary
nanostructures by electron beam lithography in such a way that access
by TEM is possible. The structures, with the carbon nanotubes embedded,
are created on the edge of a cleaved substrate and made free-standing
with an etching process. Fig. \ref{cap:SEM} shows two samples with
a free-standing section and embedded carbon nanotubes. The free-standing
structure can be freely designed by electron beam lithography.

\begin{figure}
\includegraphics[%
  width=8.5cm]{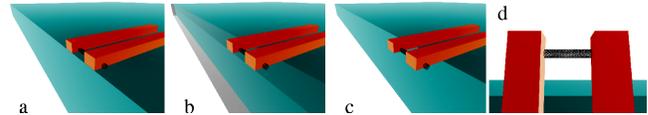}

\caption{(color online) Schematic illustration of the sample preparation procedure:
(a) The substrate is cleaved through a metallic grid which is on top
of the carbon nanotubes. (b) The sample is etched in such a way that
the structure is mainly undercut from the side, removing the shaded
volume. The resulting structure (c) reaches out across the side edge
of the substrate. Since the nanotube is still held by the metal contacts
and the substrate is no longer in the way, it is accessible for TEM
investigations (d, top view).\label{cap:Schematic-illustration}}
\end{figure}

Single-walled carbon nanotubes are grown by chemical vapour deposition
(CVD) on highly doped silicon substrates with a 200~nm Silicon dioxide
layer \cite{PailletCVDJPCB04}. A metal structure consisting of 3~nm
Cr and 110~nm Au is created by electron beam lithography on top of
the as-grown carbon nanotubes. The substrate is then cleaved through
the metal grid structure. An etching process, as illustrated in Fig.
\ref{cap:Schematic-illustration}, is used to obtain freestanding
nanotubes: The sample is etched in a 30\% KOH bath at 60°C for 7 hours.
This removes quickly the bulk Si substrate, and slowly the oxide layer.
The etch rate of the doped silicon substrate can be controlled by
biasing it with respect to the bath. Since the oxide layer initially
acts as a mask, the structure is undercut mainly from the side of
the cleaved edge. An undercut of 10~$µ$m can be achieved, and the
etching process has to be stopped just when the oxide layer is completely
removed. After the etching process, the sample is transferred into
deionized water, isopropanol, and acetone before a critical point
drying step.

Since the substrate is no longer in the way, TEM is possible on the
free-standing part of the structure on the edge of the substrate.
The carbon nanotubes are held in place by the metal structure. Before
the micro-Raman experiment, overview TEM images are obtained at low
dose and voltage (60~kV) to obtain the position and orientation of
the carbon nanotubes with respect to the metal structure (Fig. \ref{cap:SEM}).

Since the metal structure is visible in the optical microscope and
the overview images show the position of the nanotubes and their orientation,
it is possible to carry out micro-Raman experiments on an oriented
single tube. Room-temperature Raman spectra were measured using the
Ar/Kr laser lines at 488~nm (2.54~eV), 514.5~nm (2.41~eV) and
647.1~nm (1.92~eV) in the back-scattering geometry on a triple substractive
Jobin-Yvon T64000 spectrometer equipped with a liquid nitrogen cooled
charge coupled device (CCD) detector. The instrumental resolution
is 2~cm$^{-1}$. A precise positioning of the tubes under the laser
spot was monitored with a piezoelectric nano-positioner. In our experimental
configuration, the incident light polarization is along the SWNT axis
(the Z axis), and no analysis of the polarization of the scattered
light is done.

After measuring the Raman spectra, high-resolution TEM images and
diffraction images of the Raman-active nanotubes are obtained. We
obtain high-resolution images using a Philips CM200 microscope operated
at 120~kV, and we record diffraction patterns on image plates in
a Zeiss 912$\Omega$ microscope operated at 60~kV. It is operated
in the Köhler illumination condition with a condensor aperture of
20~$µ$m. The demagnified image of the condensor aperture illuminates
an area of 900~nm in diameter of the sample. We move the nanotube
into the illuminated region. Now, only this one nanotube and a small
part of the contact structure is illuminated. To exclude the contact
structure from the diffraction pattern, the selected area aperture
is used additionally to further limit the effective area to approximately
300~nm in diameter. The energy filter is set to a width 15~eV. The
diffraction pattern is exposed onto the image plate for 5 minutes
with an illumination angle of 0.1~mrad.

\begin{figure}
\includegraphics[%
  width=8.5cm]{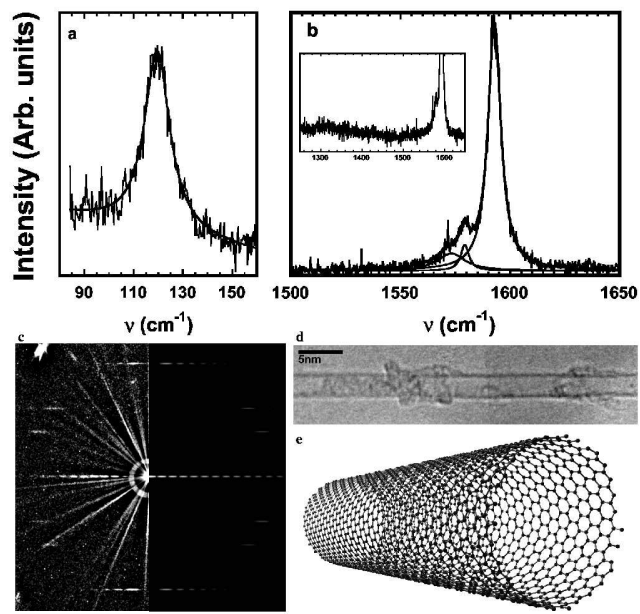}

\caption{(a) RBM and (b) TM ranges of the Raman spectrum measured on the individual
SWNT with a laser energy of 1.92~eV. (c) Diffraction pattern (left),
simulated pattern (right) and (d) high-resolution TEM image of the
same nanotube. From the diffraction pattern the nanotube can be unambiguously
identified as (27,4), while the TEM image confirms the diameter and
the presence of only one single nanotube. Image (e) is a structural
model of a (27,4) nanotube segment.\label{cap:RAM}\label{cap:DIFF}\label{cap:TEM}}
\end{figure}

Fig. \ref{cap:RAM}a and b show the Raman spectrum, and Fig. \ref{cap:TEM}c
and d show a diffraction pattern and a high-resolution TEM image obtained
from the exact same carbon nanotube.

The high-resolution image proves that the investigated tube is indeed
individual and not a bundle. From the high-resolution images alone,
the nanotube diameter can be estimated only with limited precision.
The apparent diameter depends on the defocus value, and the apparent
diameter is always smaller than the actual tube diameter \cite{DiameterMeasurementFromTEMImg02}.
A comparison with image simulations would be possible in principle
for a known defocus value, however also the point of zero defocus
can not be obtained with the desired precision.

From the diffraction pattern however, it is possible to unambiguously
identify the nanotube structure. This nanotube is identified from
the diffraction pattern alone as (27,4). By taking into account a
C-C distance of 1.42~$\pm{0.02}$ \AA, the diameter of the (27,4)
tube is 2.29~$\pm{0.03}$nm. Its chiral angle is 6.8$^{\circ}$.
Fig. \ref{cap:DIFF}c shows the diffraction pattern and a simulated
diffraction pattern of a (27,4) nanotube. We can rule out other indices
for which the simulated patterns clearly do not match the measured
one, and its diameter of 2.29~$\pm{0.03}$ nm is consistent with
the high-resolution images.

Figures \ref{cap:RAM}a and b show the RBM and TM ranges of the Raman
spectrum measured on the SWNT using a 1.92~eV excitation ($\lambda$=647.1~nm),
with the polarization in the direction of the tube axis \cite{DuesbergPRL00}.
The (27,4) tube is semiconducting, and with regards to the resonance
conditions calculated using a nearest-neighbor tight binding approach
\cite{KatauraSynthMet99}, its Raman response is expected to be enhanced
at 1.7~eV excitation. In agreement with this prediction a detectable
signal is only observed for the 1.92~eV excitation in our Raman experiments
(available energies were 1.92~eV, 2.41~eV, and 2.54~eV). As expected
for the Raman spectrum of an individual SWNT, the spectrum is featured
by a single narrow RBM. For the (27,4) tube under investigation, this
RBM has a center frequency of 119.5~cm$^{-1}$ (Fig. \ref{cap:RAM}a).
In our scattering geometry, a number of lines greater than two in
the TM bunch of the Raman spectrum is predicted due to the chiral
character of the (27,4) tube \cite{JorioPRL03}. In agreement with
this, the profile of the TM bunch, displayed in Fig. \ref{cap:RAM}b,
is well fitted by using three main Lorentzian components located at
1593~cm$^{-1}$, 1579~cm$^{-1}$ and 1573~cm$^{-1}$. As expected
for a semiconducting SWNT, no line broadening as predicted for metallic
SWNTs \cite{DresselhausAdvPhys00} in the TM bunch is observed .

In the inset of figure \ref{cap:RAM}b, the D band range is shown
in an expanded scale, and no band is observed in this region. The
strength of the D band is expected to be indicative of the amount
of defects in the nanotube. The quality of the diffraction pattern
indicates a high structural integrity of the nanotube: The structure
is periodic, i.e. the indices do not change within the illuminated
region and the amount of defects is low. The high-resolution TEM image
provides an upper limit for the amount of amorphous carbon
on the tube, since additional carbon is clearly deposited during the
TEM analysis. This is in agreement with the interpretation of the
D-band as an indicator of defects and amorphous carbon coating. The
nanotubes in our samples are completely free-standing, separated and
sufficiently clean; most environmental influences on the nanotube
spectra can be excluded. 

A huge number of experiments and modelization efforts were made to
relate the radial breathing mode (RBM) frequency to the nanotube structure.
A review and summary of various models and experiments is given in
\cite{sreich_CNTs-concepts+prop2004}. Our experimental approach gives
access to a first accurate and independent determination of a specific
SWNT diameter, at the same time with its RBM frequency. For a free-standing
individual SWNT, the RBM frequency $\nu$ is expected to be inversely
proportional to the nanotube diameter $d$, i.e. $d=\textrm{{A}}/\nu$
\cite{sreich_CNTs-concepts+prop2004}. From the results obtained on
the (27,4) SWNT, a proportionality constant of A=273$\pm$10 nm.cm$^{-1}$
is found. This surprisingly high value either indicates that the models
are underestimating the RBM frequency (see table 8.2 in Ref.\cite{sreich_CNTs-concepts+prop2004}),
or that this simple RBM vs. diameter relationship can not be extrapolated
up to the relatively large diameter of this tube. However for a precise
estimation of the RBM vs. tube diameter relationship, a diffraction
analysis of many tubes with a wide distribution of diameters is required.

As a conclusion, we have obtained Raman spectra for a precisely known
structure, determined by electron diffraction and high-resolution
TEM imaging. A measurement of the vibrational modes for a precisely
known structure provides the ultimate test for molecular-dynamics
simulations. Both the micro-Raman spectroscopy and the electron microscopic
investigation are done on a freely suspended object without an influence
from an environment, substrate or contact. We expect that this procedure,
due to the freely designable freestanding structure, can be adopted
to various nano-objects or macromolecules to combine electron microscopic
structural analysis with Raman spectroscopy and potentially other
investigations (transport, AFM) on the same object. Also, the possibility
to create arbitrary free-standing structures may facilitate novel
in-situ experiments in the TEM.

The authors acknowledge financial support by the EU project CARDECOM
and the BMBF project INKONAMI. We thank xlith.com for lithography
services. We thank P. Poncharal, A. Zahab, C. Koch, K. Hahn, M. Kelsch,
F. Phillipp and M. Rühle for support and helpful discussions.

This article has been submitted to Applied Physics Letters (http://apl.aip.org)

\bibliographystyle{apsrev}
\bibliography{Raman,TEM,diverse,books,transport}

\end{document}